\documentstyle[12pt,epsfig]{article}
\def\beq{\begin{equation}}
\def\eeq{\end{equation}}
\def\bea{\begin{eqnarray}}
\def\eea{\end{eqnarray}}

\begin{document}
\begin{flushright}
arXiv: 0901.1433 [hep-th]\\
CAS-PHYS-BHU Preprint
\end{flushright}

\vskip 1cm

\begin{center}
{ \small \bf \sf ABSOLUTE ANTICOMMUTATIVITY OF THE NILPOTENT SYMMETRIES IN THE HAMILTONIAN
FORMALISM: FREE ABELIAN 2-FORM GAUGE THEORY }

\vspace{1.5cm}

{\sf R.  P.  Malik{$^{(a,b)}$}\footnote{e-mail: malik@bhu.ac.in , rudra.prakash@hotmail.com }, 
B.  P.  Mandal{$^{(a)}$}\footnote{e-mail: bhabani.mandal@gmail.com}, 
S.  K.  Rai{$^{(a)}$}\footnote {e-mail: sumitssc@gmail.com}}\\
 {\em {$^{(a)}$}Physics Department, Centre of Advanced Studies,\\
  Banaras Hindu University, Varanasi - 221 005, India }\\
  
{\vskip 0.1cm}

{\bf and}\\

{\vskip 0.1cm}

{\it {$^{(b)}$}DST Centre for Interdisciplinary Mathematical Sciences,}\\
{\it Faculty of Science, Banaras Hindu University, Varanasi - 221 005, India}\\

\end{center}

\medskip
\bigskip

\noindent {\bf{\sf {Abstract:}}}  
The celebrated Curci-Ferrari (CF) type of restrictions are invoked to obtain the off-shell nilpotent
and {\it absolutely} anticommuting (anti-) BRST as well as (anti-) co-BRST symmetry 
transformations in the context of the Lagrangian description of the physical four (3 + 1)-dimensional 
(4D) free {\it Abelian} 2-form gauge theory. We show that the above CF type conditions,
which turn out to be the secondary constraints of the theory, remain invariant 
with respect to the time-evolution of the above 2-form gauge system in the Hamiltonian formulation.
 This time-evolution invariance (i) physically ensures the linear independence of the BRST versus anti-BRST as well as co-BRST versus anti-co-BRST symmetry transformations, and (ii) provides a logical reason behind the imposition of the above CF type restrictions in the proof of the absolute anticommutativity of the off-shell
nilpotent (anti-) BRST as well as (anti-) co-BRST symmetry transformations. \\

 
 
 \medskip
\vspace{1in}
\newpage
\section{Introduction}

The principle of local gauge invariance, in the context of the (non-) Abelian 1-form
gauge theories, has played a key role in providing a successful theoretical description of the strong, 
weak and electromagnetic interactions of nature. The existence of the first-class constraints, in the 
language of Dirac's prescription for the classification  scheme [1,2], is at the heart of the above 
(non-) Abelian 1-form
($A^{(1)} = dx^\mu A_\mu$) gauge theories which provide the cornerstones for the beautiful
edifice of the standard model of theoretical high energy physics. It is now a common 
folklore in theoretical physics that any arbitrary
$p$-form ($ p = 1, 2, 3...$) gauge theory should always be endowed with the first-class constraints. These
constraints, in fact, generate precisely the local gauge symmetry  transformations
of any specific $p$-form gauge theory in any arbitrary $D$-dimension of spacetime [1,2].

In the recent past, the 4D free Abelian 2-form ($B^{(2)} = [(dx^\mu \wedge dx^\nu)/2!] B_{\mu\nu}$) gauge field
$B_{\mu\nu}$ [3,4] has become quite popular {\it mainly} 
due to its appearance in the supergravity multiplet [5] and excited
states of the  (super)string theories [6,7]. It has
played, furthermore, a crucial role in providing a noncommutative
structure in the context of string theory [8]. We have shown,
moreover, in our earlier works [9-11], that this theory
provides a tractable field theoretical model for the Hodge theory and a model for the quasi-topological field
theory [12]. One of the most interesting observations, connected with the above theory, has come out from
its discussion in the framework of superfield formulation 
proposed in [13,14]. This has led to the existence of a Curci-Ferrari
(CF) type restriction\footnote{The appearance of the CF type restriction in the context of the 
{\it Abelian} gauge
theory is first of its kind. In fact, the superfield formulation of [13,14] 
has been applied, for the first time, to the 
Abelian {\it 2-form} gauge theory in [15]. Its application in the context of {\it 1-form} gauge theories is quite
well-known.} [15]
which happens to be the hallmark of a 4D non-Abelian 1-form gauge theory (see, e.g. [16]).

It is well-known that, for the absolute anticommutativity and existence of the off-shell nilpotent
Becchi-Rouet-Stora-Tyutin (BRST) and anti-BRST symmetry transformations, one invokes the CF restriction [16]
in the case of the description of the 4D {\it non-Abelian 1-form} gauge theory [17-20]. For the first time, 
however, it has been shown that
the replication of this CF type restriction is required in the context of the 4D {\it Abelian 2-form}
gauge theory [15] so that one could obtain (i) the absolute anticommutativity\footnote{The nilpotent (anti-) BRST symmetry transformations have been shown to be anticommuting
only up to a vector gauge transformation in the context of Abelian 2-form gauge theory (see, e.g., [10]).}
of the (anti-)BRST symmetry transformations, and (ii) an independent identity of the anti-BRST symmetry transformations (and corresponding
anti-BRST charge) [21,22]. It has been possible to obtain a set of coupled Lagrangian densities that incorporates
the above CF type restriction to demonstrate that the (anti-) BRST symmetry transformations 
(and their generators) have their own independent identity [21,22]. This CF type restriction has
also been shown to have connection with the geometrical objects called gerbes [21].

The existence of the above CF type restriction has so far been shown in the framework of (i) the superfield
formalism [15], and (ii) the Lagrangian formulations [21-23,9]. Physically, it has {\it not} been made clear as to
why this type of restrictions should be imposed in the dynamical description of the Abelian 2-form gauge theory
within the framework of BRST formalism. The purpose of our present endeavour is to answer the above query in
the framework of the Hamiltonian formulation. We demonstrate that the above CF type restrictions are the secondary constraints which are derived by requiring the time-evolution invariance  of the primary constraints
of the theory. Furthermore, we show that the above CF type restrictions remain invariant with respect to the 
time-evolution of the Abelian 2-form gauge system (within the framework of the Hamiltonian formulation). This
key result of our present investigation physically ensures the imposition of the CF type restrictions,
for the absolute anticommutativity of the (anti-) BRST and (anti-) co-BRST symmetry transformations, at any
arbitrary moment of the time-evolution.

In our earlier works (see, e.g. [9,23]), we have derived the CF type restrictions from the coupled, equivalent and
(anti-) BRST as well as (anti-) co-BRST invariant 
Lagrangian densities in two steps\footnote{First of all the Euler-Lagrange equations of motion are
derived from the coupled Lagrangian density. This is followed, then, by the subtraction and addition of
the above equations of motion.} 
by exploiting the Euler-Lagrange equations of motion. It would be economical as well as aesthetically beautiful to derive the same restrictions from a single Lagrangian
density and corresponding Hamiltonian density. We accomplish this goal in our present paper where we derive the
CF type restrictions in a single stroke
and show their time-evolution invariance from a single Hamiltonian density. The latter
property, in the context of the dynamical evolution of the Abelian 
2-form system, has been established in a convincing manner. This analysis has been performed {\it explicitly} 
so that the anticommutativity of the (anti-) BRST and (anti-) co-BRST symmetries
could be ensured at {\it each} moment of the time-evolution of our present 2-form gauge system.

Our present investigation has been motivated by the following factors. First and foremost, the time-evolution
invariance of the CF type restrictions cannot be demonstrated within the framework of {\it either} superfield
{\it or} Lagrangian formulation. Thus, it is essential for us to describe the Abelian 2-form gauge
system within the framework of the Hamiltonian approach. Second, for aesthetic reasons, it is always desirable
to obtain the CF type restrictions from a single Lagrangian density (and corresponding Hamiltonian density).
We have accomplished this goal in our present endeavour. Finally, our present attempt is a modest step in
the direction to provide the physical reasons behind the appearance of the CF type restrictions in the 
context of the higher $p$-form ($p > 2$) gauge theories within the framework of BRST formalism. Thus, our present study might have relevance in the description 
of the higher-form fields (associated with string and other extended objects).

The outline of our present paper is as follows. To set up the conventions and notations, we briefly mention in Sec. 2, the (anti-)BRST symmetries in the Lagrangian formulation. Our Sec. 3 is devoted to the discussion of the time-evolution invariance of the CF-type restriction that is invoked for the proof of anticommutativity of the off-shell nilpotent (anti-) BRST symmetries in the Hamiltonian formulation. For the paper to be self-contained,
in Sec. 4, we provide a brief synopsis of the (anti-) co-BRST symmetries within the framework of  Lagrangian
formalism. Our Sec. 5 deals with the time-evolution invariance of the CF type restriction, in the framework of Hamiltonian formulation, that is required in the proof of the absolute anticommutativity of the off-shell nilpotent (anti-)co-BRST symmetry transformations. Finally, in Sec. 6, we make some concluding remarks and point out a few new directions for future investigations.

\section{Preliminaries: Off-shell Nilpotent (Anti-) BRST Symmetries in Lagrangian Formulation}

We begin with the following Lagrangian densities for the 4D free abelian 2-form gauge theory {\footnote{We adopt
here the conventions and notations such that the flat 4D Minkowski metric $\eta _{\mu \nu }$ is with signature (+1, -1, -1, -1). The 4D totally antisymmetric Levi-Civita tensor is chosen to obey 
$\varepsilon _{\mu \nu \eta \kappa }
\varepsilon^{\mu \nu \eta \kappa} = - 4! $, $\varepsilon_{\mu \nu \eta \kappa }
\varepsilon ^{\mu \nu \eta \xi }=-3!\delta ^{\xi}_\kappa$, etc., and 
$\varepsilon _{0123}= +1=-\varepsilon ^{0123}$. The 3D Levi-Civita tensor 
is defined as: $\varepsilon _{0ijk}=\epsilon _{ijk}$. Here the Greek indices
 $\mu, \nu, \eta, \kappa.....= 0, 1, 2, 3 $ correspond to the spacetime directions 
of the 4D Minkowski spacetime manifold and 
Latin indices $i, j, k....= 1, 2, 3 $ stand for space directions only.}}
within the framework of the  BRST formalism (see, e.g, \cite{guma})
\begin{eqnarray}
{\cal{L}}^{(1)} &=& \frac{1}{12}\; H^{\mu \nu \kappa }\; H_{\mu \nu \kappa } + B^{\mu} \left(\partial^{\nu }B_{\nu\mu }+ \frac{1}{2}\partial _{\mu}\varphi_1 \right ) - 
\frac{1}{2} B^{\mu}B_{\mu} + \partial_{\mu} \bar 
{\beta } \partial^{\mu} \beta \nonumber\\ 
 &+& \left(\partial _{\mu}\bar{C}_\nu - \partial _{\nu}\bar{C}_\mu \right )
\;\partial^{\mu} C^\nu + \left (\partial \cdot C - \lambda \right ) \; \rho + \left (\partial \cdot \bar{C} + \rho \right )
\; \lambda,  \label{l1}
\end{eqnarray}
\begin{eqnarray}
{\cal{L}}^{(2)} & = & \frac{1}{12} H^{\mu \nu \kappa } H_{\mu \nu \kappa } + 
\bar{B}^{\mu} \left (\partial^{
\nu } B_{\nu\mu } - \frac{1}{2} \partial _{\mu}\varphi_1 \right ) - \frac{1}{2} \bar{B}^{\mu} \bar{B}_{\mu} + \partial_{\mu} 
\bar{\beta }\partial^{\mu} \beta \nonumber\\ 
 &+& \left (\partial _{\mu} \bar{C}_\nu - \partial _{\nu} \bar{C}_\mu \right )\;
\partial^{\mu} C^\nu + \left(\partial \cdot C - \lambda \right )\; \rho + \left (\partial \cdot \bar{C} + \rho \right )
\;\lambda,  \label{l2}
\end{eqnarray}
where the totally antisymmetric curvature tensor $ H_{\mu \nu \kappa }= \partial_\mu B_{\nu 
\kappa }+\partial_{\nu}B_{\kappa \mu }+\partial_{\kappa}B_{\mu \nu } $ is derived from the 
3-form $ H^{(3)}= dB^{(2)}=[(dx^{\mu}\wedge dx^{\nu}\wedge dx^{\kappa})/3!]H_{\mu \nu 
\kappa } $ constructed with the help of the exterior derivative $ d=dx^{\mu}\partial _{\mu}$ 
(with $d^2=0$) and the Abelian 2-form connection $ B^{(2)}=[(dx^{\mu}\wedge dx^{\nu})/2!] B_{\mu\nu} $ which  
defines the antisymmetric $(B_{\mu\nu} = -B_{\nu\mu})$ gauge potential $B_{\mu\nu}$ of the present Abelian 2-form gauge theory.

The BRST invariance in the theory requires the fermionic
$(C_{\mu}\bar{C}_{\nu}+\bar{C}_{\nu}C_{\mu}=0, C_{\mu}^2=0,\bar{C}_\mu^2=0,$ etc.)  Lorentz vector
(anti-) ghost $(\bar{C}_\mu)$ $C_\mu $ fields, fermionic ($\rho ^2=\lambda ^2=0,\;\rho \lambda + \lambda \rho =0$ ) auxiliary (anti-) ghost fields $\rho $ and
$ \lambda$ and bosonic ($\beta^2\neq 0,\;\bar{\beta}^2\neq0,\;\beta\bar{\beta}=\bar{\beta}\beta )$ (anti-) ghost fields $(\bar{\beta })\beta$ . In the above, 
$ B_{\mu} $ and $\bar{B}_\mu$ are the Nakanishi-Lautrup type of auxiliary fields that are invoked for the 
linearization of the gauge fixing terms $[\frac{1}{2}{(\partial^{\nu 
}B_{\nu\mu }+ \frac{1}{2}\partial _{\mu}\varphi_1 )}^2]$ and $[\frac{1}{2}{(\partial^{\nu 
}B_{\nu\mu }- \frac{1}{2}\partial _{\mu}\varphi_1 )}^2]$ where $\varphi_1$ is the massless 
($\Box{\varphi _1}=0$) scalar 
field required for the stage-one reducibility in the theory. The gauge-fixing term 
($\partial^{\nu 
}B_{\nu\mu }$) owes its origin to the co-exterior derivative $\delta =-*d*$ because $\delta B^{(2)}=(\partial^{\nu 
}B_{\nu\mu }) dx^{\mu}$ where $*$  is the Hodge duality operator on the 4D spacetime manifold.

The following off-shell nilpotent $(s_b^2=0)$ BRST transformations ($s_b$)
\begin{eqnarray}
& s_b B_{\mu\nu} = -(\partial_{\mu}C_{\nu}-\partial _{\nu}C_{\mu}),\; \quad s_b C_{\mu}= -\partial_{\mu}\beta,
\quad \; s_b \bar{C}_\mu = -B_\mu, & \nonumber\\
& s_b\varphi_1 = - 2 \lambda, \; \quad s_b\bar{\beta} = -\rho, \quad \; s_b\bar{B}_\mu = - \partial_\mu\lambda, 
\quad \; s_b [\rho,\lambda,\beta,B_{\mu},H_{\mu \nu \kappa }] = 0, &
\end{eqnarray}
and the off-shell nilpotent ($s_{ab}^2=0$) anti-BRST transformations ($s_{ab}$)
\begin{eqnarray}
& s_{ab} B_{\mu\nu} = - (\partial_\mu\bar{C}_\nu - \partial_\nu\bar{C}_\mu ), \; \quad s_{ab}\bar
{C}_{\mu}= -
\partial_{\mu}\bar{\beta}, \; \quad s_{ab} C_\mu = \bar{B}_\mu,  &\nonumber\\
&s_{ab}\varphi_1  = - 2 \rho,  \; \quad s_{ab} \beta = - \lambda, \; \quad s_{ab}B_\mu = 
\partial_\mu\rho, \; \quad
s_{ab}[\rho,\lambda,\bar{\beta},\bar{B}_{\mu},H_{\mu \nu \kappa }] = 0, & 
\end{eqnarray}
are

(i) the symmetry transformations for the Lagrangian densities (\ref{l1}) and (\ref{l2}) \cite{guma}, and 
 
(ii) absolutely anticommuting ($s_b s_{ab} + s_{ab} s_b = 0$) in nature because
their absolute anticommutativity property (e.g. $\{s_b,s_{ab}\}B_{\mu\nu}=0$) is ensured due 
to the following Curci-Ferrari (CF) type of restriction
\begin{equation}
B_\mu -\bar{B}_\mu-\partial_\mu\varphi_1 =0. \label {cf}
\end{equation} 
The above condition emerges from (\ref{l1}) and (\ref{l2}) due to the equations of motion $ B_\mu=\partial^\nu B_{\nu 
\mu}+\frac{1}{2}\partial_\mu\varphi_1,\quad \bar{B}_\mu = \partial^\nu B_{\nu 
\mu}-\frac{1}{2}\partial_\mu\varphi_1 $ .
The key points that ought to be noted,  at this stage, are as follows. First, it can be seen 
that 
the CF type restriction (\ref {cf}) is derived in two steps from the Lagrangian densities (\ref
{l1}) and (\ref{l2}). Second, unlike in the context of the 4D non-Abelian 1-form gauge theory where 
the (anti-) ghosts fields also participate in the CF condition \cite{cufe}, for the Abelian 2-form gauge theory only the bosonic fields contribute to its existence. Finally, the time evolution invariance of the CF 
type condition (\ref{cf}) is not guaranteed in the Lagrangian description of the free 4D Abelian 2-form gauge theory. Thus, the logical reason behind the imposition of the CF type restriction (\ref{cf}), for the above anticommutativity property, is not clear within the framework of the Lagrangian formalism. This is why, in the next section, we resort to the Hamiltonian formalism.\\

\section{Time-Evolution Invariance of the Curci-Ferrari Type Condition: Hamiltonian Approach}

It can be noted that the ghost part of the Lagrangian densities (\ref{l1}) and (\ref{l2}) is same.     The corresponding Hamiltonian density $({\cal{H}}_{(g)} )$ can be expressed as 
\begin{eqnarray}
{\cal{H}}_{(g)} & = & \Pi^{(\beta)} \;\Pi^{(\bar{\beta})} + \Pi _{i}^{(c)}\;\Pi_{i}^{(\bar{c})} +\Pi_{i}^{(\bar{c})}\;(\partial _iC_0) + \Pi_{i}^{(\bar{c})}\;(\partial_i\bar{C}_0) 
+ \partial_i \bar{\beta} \partial_i{\beta} \nonumber\\
& - &\left (\partial _i\bar{C}_j-\partial_j\bar{C}_i\right) \;\partial_i C_j + (\partial_i C_i)
\;\Pi_0^{(c_0)} + (\partial_i \bar{C}_i)\; \Pi_0^{(\bar{c}_0)} + 2 \Pi_0^{(\bar{c}_0)}\; \Pi_0^{(c_0)},
\label{hg}
\end{eqnarray}
where the canonical momenta, corresponding to the (anti-) ghost fields, are:
\begin{eqnarray}
\Pi^{(\beta)} & \equiv & 
\frac{\partial{\cal{L}}^{(1,2)}}{\partial(\partial _0\beta)} \;=\dot{\bar{\beta}},\;\;\quad \;
\Pi^{({\bar{\beta}})} \equiv \frac{\partial{\cal{L}}^{(1,2)}}{\partial(\partial _0 \bar{\beta})} =
\dot{\beta}, \nonumber\\
\Pi_0^{(c_0)} &\equiv&\frac{\partial{\cal{L}}^{(1,2)}}{\partial (\partial _0 C_0)}=\rho, \quad \;\Pi_0^
{(\bar{c}_0)} \equiv \frac{\partial{\cal{L}}^{(1,2)}}{\partial (\partial _0\bar{C}_0)}=\lambda, \nonumber \\
\Pi_{i}^{(\bar{c})} &\equiv& \frac{\partial{\cal{L}}^{(1,2)}} {\partial (\partial ^0 \bar C^i)}
= (\partial _0 C_i-\partial_i C_0),  \nonumber\\
\Pi_{i}^{(c)}&\equiv& \frac{\partial{\cal{L}}^{(1,2)}}{ \partial (\partial ^0 {C}^i)}
= - (\partial _0\bar{C}_i-\partial_i \bar{C}_0).\label{cm}
\end{eqnarray}
It is worthwhile to mention that, in the operation of the derivative w.r.t the fermionic ghost fields, we have adopted the convention of the left derivative.

The following Heisenberg equations of motion for the generic field $\Psi$
\begin{eqnarray}
\dot{\Psi } = \pm \;i \left [ \Psi , H_{(g)} \right ],\;\quad \quad H_{(g)} = 
{\displaystyle \int} d^3 x \;{\cal{H}}_{(g)},\;\quad \quad 
\dot{\Psi }=\frac{\partial \Psi}{\partial  t},  \label {hb}
\end{eqnarray}
(where [(+)-] signs correspond to the (fermionic)bosonic nature of the generic field $\Psi$) lead to the dynamical equations of motion for momenta as well as basic fields. It can be 
checked that the Euler Lagrange equations of motion
\begin{eqnarray}
&\Box{\beta} =\Box{\bar{\beta}} =0,\;
\quad \quad\Box{\bar{C}_0}=-\partial_0\rho, 
\quad\quad \Box{C_0}=\partial_0 \lambda, & \nonumber\\
&\Box{\bar{C}_i} = -\partial_i\rho, 
\quad\quad \Box{C_i}=\partial_i\lambda,\;\quad \lambda =\frac{1}{2}(\partial\cdot C),
\;\quad \rho =-\frac{1}{2}(\partial\cdot\bar{C}),& \label{ag}
\end{eqnarray}
for the (anti-)ghost fields, derived from the Lagrangian densities ${\cal{L}}^{(1,2)}$, 
also emerge from equation (\ref{hb}) when $\Psi =\Pi^{(\beta)}, \Pi^{({\bar{\beta}})}, \Pi^{(c_0)}, \Pi^{(\bar{c}_0)}, \Pi_{i}^{(\bar{c})}, \Pi_{i}^{(c)}$. On the other hand, for $\Psi = \beta, \bar{\beta}, 
C_0, \bar{C}_0, C_i, \bar{C}_i$, we obtain the definition of the canonical momenta (\ref{cm}). In our Appendix A, these explicit computations are illustrated in a detailed fashion.

The non-ghost parts of the Lagrangian density (\ref{l1}) and (\ref{l2}) lead to the following 
pair of the canonical Hamiltonian densities in terms of canonical momenta and fields:
\begin{eqnarray}
{\cal{H}}^{(1)}_{(b)} &=&({\Pi _{ij}})^2 + 2\; ({\Pi _{\varphi _1}^{(1)}})^2  -\frac{1}{2} \;({\Pi 
_{0i}^{(1)}})^2 - 2\;\Pi _{ij}\;
 (\partial _i B_{j0}) + \frac{1}{2}\;(\Pi _{0i}^{(1)})\;\partial _i \varphi _1 \nonumber\\
&-&(\Pi_{0j}^{(1)})\; \partial
 _i B_{ij} - 2\;(\Pi _{\varphi _1}^{(1)}) \;\partial _i B_{0i} 
+ \frac{1}{12}\; H_{ijk}\;H_{ijk},  \label{hb1}
\end{eqnarray}
\begin{eqnarray}
{\cal{H}}^{(2)}_{(b)} &=&{
(\Pi _{ij}})^2 + 2\;({\Pi _{\varphi _1}^{(2)}})^2  -\frac{1}{2}\;({\Pi _{0i}^{(2)}})^2 - 2\;\Pi _{ij}
\;(\partial _i B_{j0}) - \frac{1}{2}\;(\Pi _{0i}^{(2)})\;\partial _i \varphi _1 \nonumber\\
&-&(\Pi_{0j}^{(2)}) \; \partial
 _i B_{ij} - 2\; (\Pi _{\varphi _1}^{(2)})\; \partial _i B_{0i} 
+ \frac{1}{12}\; H_{ijk}\; H_{ijk},  \label{hb2}
\end{eqnarray}
where the canonical momenta are defined as follows
\begin{eqnarray}
\Pi_{ij}&\equiv & \frac{\partial {\cal{L}}^{(1,2)}}{\partial (\partial ^0B^{ij})} =\frac{1}{2} 
H_{0ij},\nonumber\\
\Pi_{0i}^{(1)} &\equiv &\frac{\partial {\cal{L}}^{(1)}}{\partial (\partial ^0B^{0i})}=  B_i, \;\quad 
\Pi_{0i}^{(2)} \equiv \frac{\partial {\cal{L}}^{(2)}}{\partial (\partial ^0B^{0i})}=  \bar{B}_i, 
\nonumber\\
\Pi_{\varphi _1}^{(1)}&\equiv & \frac{\partial {\cal{L}}^{(1)}}{\partial (\partial _0\varphi _1)} =
\frac{B_0}{2}, \; \quad \Pi_{\varphi _1}^{(2)}\equiv \frac{\partial {\cal{L}}^{(2)}}{\partial (
\partial _0\varphi _1)} = -\frac{\bar{B}_0}{2}.
\end{eqnarray}
Exploiting the appropriate form of the Heisenberg equation (\ref{hb}) with the Hamiltonian 
densities (\ref{hb1}) and (\ref{hb2}) and using the following canonical brackets \footnote{All the rest of the brackets are zero.} (with $\hbar$
 = c=1)
\begin{eqnarray}
\left [B_{ij}({\bf{x}},t) ,\Pi_{kl}({\bf{y}},t)\right ]&=& \frac{i}{2} (\delta _{ik} \delta _{jl}-\delta_
{il} \delta _{jk}) \delta ^{(3)}({\bf{x}}-{\bf{y}}), \nonumber\\
\left [B_{0i}({\bf{x}},t) ,\Pi _{0j}^{(1,2)}({\bf{y}},t)\right ]&=& -i  \delta _{ij} \delta ^{(3)}({\bf{x}}-{\bf
{y}}), \nonumber\\
\left[\varphi _1({\bf{x}},t) ,\Pi_{\varphi_1}^{(1,2)}({\bf{y}},t)\right ] &=& i \delta ^{(3)}(
{\bf{x}}-{\bf{y}}), \label{cb}
\end{eqnarray} 
it can be checked that the Hamiltonian densities (\ref{hb1}) and (\ref{hb2}) produce all the 
Euler Lagrange equations of motion derived from the Lagrangian densities (\ref{l1}) and      (\ref{l2}). These derivations are clearly illustrated in our Appendix B. It will be noted that the CF condition (\ref{cf}) is still not derivable from a single Hamiltonian density (\ref{hb1}) and/or (\ref{hb2}). The CF condition (\ref{cf}) can be derived in one stroke, however. Towards, this goal in mind, we define the following Lagrangian density\footnote{It will be noted that the other linearly independent combination $\frac{1}{2} [{\cal L}^{(1)} - {\cal L}^{(2)}]$
is {\it not} interesting because the  kinetic term of the gauge field and the ghost part of the Lagrangian
densities cancel out in this combination. Thus, this combination
 is {\it not} useful from the point of view of our present discussions.}
that is constructed from (\ref{l1}) and (\ref{l2}), namely; 
\begin{eqnarray}
{\cal{L}}^{(3)} &=& \frac{1}{2} \left({\cal{L}}^{(1)}+{\cal{L}}^{(2)}\right)
\quad \equiv \quad \frac{1}{12}H^{\mu \nu \kappa }\; H_{\mu \nu \kappa } + \frac{1}{2}
\left(B^\mu+\bar{B}^\mu\right )\;\partial 
^\nu B_{\nu \mu }\nonumber\\ 
&+&\frac{1}{4}\left(B^\mu-\bar{B}^\mu\right)\;\partial_\mu\varphi
 _1- \frac{1}{4}\left(B\cdot B+\bar{B}\cdot{\bar{B}}\right) + {\cal{L}}_{(g)}, \label{l3}
\end{eqnarray}
where 
\begin{equation}
{\cal{L}}_{(g)}= \partial_{\mu} \bar 
{\beta }\partial^{\mu} \beta 
 + \left(\partial _{\mu}\bar{C}_\nu - \partial _{\nu}\bar{C}_\mu \right)\;
\partial ^{\mu}C^\nu + (\partial \cdot C -\lambda )\;\rho +(\partial \cdot \bar{C}+\rho )
\;\lambda,  \label{lg}
\end{equation}
is the ghost part of the Lagrangian densities (\ref{l1}) and/or (\ref{l2}).
It can be checked that, even from the Lagrangian density (\ref{l3}), the CF type of restriction (\ref{cf}) can be derived only in two steps. To obtain the same condition (i.e. (\ref{cf}))
in a single stroke, one has to redefine the following pair of auxiliary fields:
\begin{equation}
b_\mu = \frac{1}{2} (B_\mu +\bar{B}_\mu ), \; \quad \quad \bar{b}_\mu =\frac{1}{2} (B_\mu -\bar{B}_\mu). \label{af1}
\end{equation}
As a result of the above re-definitions, it can be shown that the following equality 
\begin{equation}
B\cdot B +\bar{B} \cdot \bar{B} = 2 \;(  b\cdot b +\bar{b}\cdot \bar{b}  ),
\end{equation}
leads to a different looking form of (\ref{l3}), namely,
\begin{equation}
{\cal{L}}^{(3)} = \frac{1}{12}H^{\mu \nu \kappa }\;H_{\mu \nu \kappa } + b^\mu\;
\partial 
^\nu B_{\nu \mu } + \frac{1}{2}\bar{b}^\mu \;\partial_\mu\varphi _1
- \frac{1}{2}(b\cdot b+\bar{b}\cdot{\bar{b}}) +{\cal{L}}_{(g)}.\label{l3m}
\end{equation}
From the  very outset, it is clear that
\begin{equation}
\Pi _{(b)}^\mu \equiv \frac{\partial{\cal{L}}^{(3)}}{\partial (\partial_0 b_\mu)} = 0, \;
 \quad \quad  \Pi _{(\bar{b})}^\mu \equiv 
\frac {\partial{\cal{L}}^{(3)}}{\partial (\partial_0 \bar{b}_\mu)} = 0, \label{pc}
\end{equation}
are the primary constraints on the theory.
The canonical Hamiltonian density, derived from the Lagrangian density (\ref{l3m}), is
\begin{eqnarray}
{\cal{H}}_{(b,\bar{b})}^{(3)} &=& \Pi _{ij}^2+ 2\;\Pi _{\varphi _1}^2  -\frac{1}{2}\;{\Pi 
_{0i}}^2 
- 2\;\Pi _{ij}\;\partial _j B_{0i}
 - \Pi_{0j}\;\partial_i B_{ij} -b_0\;\partial _i B_{0i} \nonumber \\
&+& \frac{1}{2}\; \bar{b}_i \;\partial _i \varphi _1 + \frac{1}{2}\;\left(b_0b_0 - \bar{b}_i \bar{b}_i\right ) +\frac
{1}{12}\; H_{ijk} \;H_{ijk}+ {\cal{H}}_{(g)}, \label{ch}
\end{eqnarray}
where the other canonical momenta, besides (\ref{pc}) for the Lagarngian density (\ref{l3m}), are
\begin{equation}
\Pi_{\varphi_1}= \frac{\bar{b}_0}{2}, \;\qquad \Pi_{0i}=b_i, \;\qquad \Pi_{ij} =\frac{1}{2} H_{0ij}.
\end{equation}
It is trivial to note that the auxiliary fields $b_0$ and $\bar{b}_i$ appear in the above Hamiltonian density but corresponding momenta are not present. The latter happen to be the primary constraints on the theory as is evident from (\ref{pc}). These can be added to the canonical Hamiltonian (\ref{ch}) in the following manner (see, e.g, \cite{pam,sun})
\begin{eqnarray}
{\cal{H}}_{(b,\bar{b})}^{(3)} &=& \Pi_0 ^{(b_0)} \;\partial _0 b_0 - \Pi_i ^{(\bar{b})}\;\partial _0\bar{b}_i+ \Pi _{ij}^2+ 2\;\Pi _{\varphi _1}^2  -\frac{1}{2}\;{\Pi 
_{0i}}^2 
- 2\;\Pi _{ij}\;\partial _j B_{0i}
 -\Pi_{0j}\;\partial_i B_{ij}\nonumber \\
& -& b_0\;\partial _i B_{0i}
+ \frac{1}{2}\; \bar{b}_i\;\partial _i \varphi _1 +\frac{1}{2}\;\left(b_0b_0 - \bar{b}_i \bar{b}_i\right) +\frac
{1}{12}\; H_{ijk} \;H_{ijk}+ {\cal{H}}_{(g)}, \label{chf}
\end{eqnarray}
where ${\cal{H}}_{(g)}$ is the usual ghost part of the Hamiltonian (cf. (\ref {hg})) and $\Pi_0 ^{(b_0)}$ $ \Pi_i ^{(\bar{b})}$ are the momenta corresponding to the co-ordinate fields $b_0$ and $\bar{b}_i$ (cf. (\ref{pc})). It will be noted that one can also add $\Pi_0^{(\bar{b})}\;\partial_0\bar{b}_0 - \Pi_i^{(b)}\;\partial_0b_i $ in the Hamiltonian density (\ref{chf}) but these do not play any significant role as:  $\dot{\Pi}_0^{(\bar{b})}=0,\;\dot{\Pi}_i^{(b)}=0,\;\dot{\bar{b}}_0=\dot{b}_0,\;\dot{b}_i=\dot{b}_i$.

With the help of the  canonical brackets (\ref {cb}) and the following
\begin{eqnarray}
\big [ \;b_0({\b{x}},t) ,\Pi_0^{(b_0)}({\bf{y}},t) \;\big ] &=& i \delta ^{(3)}({\bf{x}}-{\bf{y}}),\nonumber\\
\big [ \bar b_i({\bf{x}},t) ,\Pi_j^{(\bar b)}({\bf{y}},t) \big ] 
&=& -i \delta _{ij}\delta ^{(3)}({\bf{x}}-{\bf{y}}),
\end{eqnarray} 
we obtain the equations of motion as given below 
\begin{eqnarray}
&\bar{b}_\mu = \frac{1}{2}\partial _\mu \varphi_1, \;\qquad \partial \cdot{\bar{b}}=0,\; \qquad \Box\varphi_1 =0,&\nonumber\\
&b_\mu = \partial ^\nu B_{\nu\mu},\; \qquad \partial \cdot b=0,\; \qquad \Pi_{ij}=\frac{1}{2}H_{0ij}, &\nonumber\\
&\partial _\mu H^{\mu \nu \kappa } + (\partial ^\nu b^\kappa  - \partial ^\kappa  b^\nu )=0.&
\end{eqnarray}
It is worth emphasizing that $\bar{b}_\mu=\frac{1}{2}\partial _\mu\varphi_1 $  (which leads to 
$B_\mu-\bar{B}_\mu -\partial _\mu\varphi_1 =0 $ ) and $b_\mu=\partial ^\nu B_{\nu \mu} $ (i.e 
$B_\mu +\bar{B}_\mu =2\partial ^\nu B_{\nu\mu}$ ) are obtained from the Hamiltonian density 
${\cal{H}}_{(b,\bar{b})}^{(3)}$ by exploiting the Heisenberg equation of motion
$\dot{\Pi}_0^{(b_0)} = 0,\;\dot{\Pi}_i^{(\bar{b})} = 0,\;\dot{\varphi_1}= - i \left [ \varphi _1 , H_{(b,\bar{b})}^{(3)} \right]$,$\;$ and $ 
\dot{B}_{0i} = -i \left [ B_{0i} , H_{(b,\bar{b})}^{(3)}\right]$
where $H_{(b,\bar{b})}^{(3)} =\int d^3x \;{\cal{H}}_{(b,\bar{b})}^{(3)}$. This establishes the fact that 
\begin{eqnarray}
\bar{b}_\mu &=& \frac{1}{2}\partial _\mu\varphi_1\quad \Rightarrow \quad B_\mu-\bar{B}_\mu-\partial _\mu\varphi_1 =0, \nonumber\\
b_\mu &=& \partial ^\nu B_{\nu\mu} \quad \Rightarrow \quad B_\mu+ \bar{B}_\mu -2\partial ^\nu B_{\nu\mu}=0,
\label{cons}
\end{eqnarray}
are the secondary constraints on the theory.
%{\footnote{It should be noted that the time-evolution invariance of the primary constraints (i.e  $\dot{\Pi}_0^{(b_0)} = 0 \Rightarrow b_0=\partial^iB_{i0}$ and $\dot{\Pi}_i^{(\bar{b})} = 0 \Rightarrow \bar{b}_i=\frac{1}{2}\partial_i\varphi _1$) leads to only a part of the constraints (\ref {cons}). The full set of constraints (\ref {cons}) are obtained with the time-evolution of fields $\varphi _1$, $B_{0i}$ together with $\dot{\Pi}_i^{(\bar{b})} = 0, \dot{\Pi}_0^{(b_0)} = 0$.}}

The time-evolution invariance of the above constraints (i.e. $ \bar{b}_\mu =\frac{1}{2}\partial _\mu\varphi_1, b_\mu=\partial ^\nu B_{\nu\mu}$ ) can be seen to be true as:
\begin{eqnarray}
\left [2\bar{b}_0 -\partial _0\varphi_1  , H_{(b,\bar{b})}^{(3)} \right ]& = & 0,\; \quad
\left [2\bar{b}_i - \partial_ i\varphi_1 , H_{(b,\bar{b})}^{(3)} \right ] =0,\; \nonumber\\
\left [b_0 -\partial _i B_{0i} , H_{(b,\bar{b})}^{(3)} \right ]& =& 0,\; \quad
\left [b_i - \partial _0 B_{0i} - \partial_j B_{ij} , H_{(b,\bar{b})}^{(3)}\right ] = 0.
\end{eqnarray}
This establishes the time-evolution invariance of the CF type conditions which are invoked in the proof of the anticommutativity of the nilpotent (anti-) BRST symmetries.

\section{(Anti-) Dual BRST Symmetries in Lagrangian Formulation: A Brief Sketch}

The kinetic term ($ \frac{1}{12}H^{\mu \nu \kappa }\;H_{\mu \nu \kappa }$) of the Lagrangian 
densities (\ref{l1}) and (\ref{l2}) can be linearized by introducing the Nakanishi-Lautrup 
type of auxiliary fields ${\cal{B}}_\mu$ and $\bar{\cal{B}}_\mu$ and a massless ($\Box\varphi _2=0$) field 
$\varphi _2$ as given below (see, e.g. \cite{guma}):
\begin{eqnarray}
{\cal{L}}^{(4)} &=& \frac{1}{2}\;{\cal{B}}^\mu {\cal{B}}_\mu- {\cal{B}}^\mu \left ( \frac{1}{2}
\varepsilon 
_{\mu \nu \eta \kappa }\partial ^\nu B^{\eta \kappa } + \frac{1}{2}\partial _\mu \varphi _2
\right )+ B^
\mu \left (\partial^\nu B_{\nu \mu } +\frac{1}{2}\partial_\mu \varphi _1 \right )\nonumber\\
&-&\frac{1}{2}\;B^\mu B_\mu + {\cal{L}}_{(g)}, \label{l4}
\end{eqnarray}
\begin{eqnarray}
{\cal{L}}^{(5)} &=& \frac{1}{2}\;\bar{{\cal{B}}}^\mu \bar{{\cal{B}}}_\mu - \bar{{\cal{B}}}^\mu \left ( \frac{1}{2}
\varepsilon 
_{\mu \nu \eta \kappa }\partial ^\nu B^{\eta \kappa } - \frac{1}{2}\partial _\mu \varphi _2
\right )+\bar{B}^
\mu \left (\partial^\nu B_{\nu \mu } -\frac{1}{2}\partial_\mu \varphi _1 \right )\nonumber\\
&-&\frac{1}{2}\;\bar{B}^\mu \bar{B}_\mu +{\cal{L}}_{(g)}, \label{l5}
\end{eqnarray}
where ${\cal{L}}_{(g)}$ is same as the ghost part of the Lagrangian densities (\ref{l1}) and (
\ref{l2}) and $ \varphi _2 ,\; {\cal{B}}^\mu$  and $ \bar{{\cal{B}}}^\mu$ satisfy the following equations of motion 
\begin{equation}
{\Box{\varphi }}_2 = 0,\; \quad {\cal{B}}_\mu =\frac{1}{2}
\varepsilon_{\mu \nu \eta \kappa }\partial ^\nu B^{\eta \kappa } + \frac{1}{2}\partial _\mu \varphi_2,
\; \quad \;
\bar{{\cal{B}}}_\mu = \frac{1}{2} \varepsilon 
_{\mu \nu \eta \kappa }\partial ^\nu B^{\eta \kappa } - \frac{1}{2}\partial _\mu \varphi _2,
\end{equation}
which lead to a set of  CF type restrictions 
\begin{equation}
{\cal{B}}_\mu + \bar{{\cal{B}}}_\mu = \varepsilon 
_{\mu \nu \eta \kappa }\partial ^\nu B^{\eta \kappa },\; \quad \quad {\cal{B}}_\mu - \bar{{\cal
{B}}}_\mu = \partial_\mu \varphi _2. \label{cf6}
\end{equation}
It is clear that the derivation of (\ref {cf6}), from (\ref{l4}) and (\ref{l5}), is a two step 
process.

It has been demonstrated in our earlier works (see, e.g. \cite{guma}) that the Lagrangian densities (\ref{l4}) 
and (\ref{l5}) are endowed with (anti-) BRST symmetry transformation as well as absolutely 
anticommuting ($ s_d s_{ad} + s_{ad}s_d$=0) (anti-)co-BRST symmetry transformations $(s_{(a)d})$. 
The latter symmetry transformations are \cite{guma,hama}
\begin{eqnarray}
&s_d B_{\mu\nu}= -\varepsilon  _{\mu \nu \eta \kappa }\partial^\eta\bar{C}^\kappa,\; \quad  \quad 
s_d\bar{C}_\mu = -\partial_\mu \bar{\beta},\; \quad  \quad s_d C_\mu = -{\cal{B}}_\mu, & 
\nonumber\\
& s_d\varphi _2 = 2\rho,\; \quad  \quad s_d \beta = -\lambda,\; \quad  \quad s_d \left[\rho ,
\lambda ,\bar{\beta} ,\varphi_1 ,{\cal{B}}_\mu ,B_\mu, \partial^\nu B_{\nu\mu}\right ] = 0, &
\end{eqnarray}
\begin{eqnarray}
&s_{ad} B_{\mu\nu}= -\varepsilon  _{\mu \nu \eta \kappa }\partial^\eta C^\kappa, \quad  \quad 
s_{ad}C_\mu = \partial_\mu \beta,\; \quad  \quad s_{ad}\bar{C}_\mu = \bar{{\cal{B}}}_\mu, & 
\nonumber\\
& s_{ad}\varphi _2 = 2\lambda, \; \quad  \quad s_{ad}\bar{\beta} = -\rho,\; \quad  \quad s_{ad} 
\left[\rho ,\lambda ,\beta ,\varphi_1 ,\bar{{\cal{B}}}_\mu ,\bar{B}_\mu, \partial^\nu B_{
\nu\mu}\right ] =0, &
\end{eqnarray}
where

 (i) the off-shell nilpotent ($s_{(a)d}^2$) (anti-)co-BRST  symmetry transformations 
$(s_{(a)d})$ leave 

the gauge fixing terms ($\partial^\nu B_{\nu\mu}\pm \frac{1}{2} \partial_\mu\varphi _1$ ) invariant,

(ii) the co-BRST symmetry transformations ($s_d$) absolutely anticommute with the 

anti-co-BRST symmetry transformations  ($s_{ad}$) (i.e $s_ds_{ad}+s_{ad}s_d=0)$, and

(iii) the absolute anticommutativity property is ensured if and only if the condition 

${\cal{B}}_\mu - \bar{{\cal{B}}}_\mu - \partial_\mu\varphi _2=0$ (cf. (30) is imposed 
(i.e $ \{s_d, s_{ad} \} B_{\mu\nu} = 0$).\\
The time-evolution invariance of the above condition cannot be proven within the framework of 
the Lagrangian description. Thus, in the next section, we discuss the time-evolution 
invariance of ${\cal{B}}_\mu - \bar{{\cal{B}}}_\mu - \partial_\mu\varphi _2=0$ in the 
framework of Hamiltonian formalism.\\

\section{Anticommutativity of the (Anti-) Dual BRST symmetries: Hamiltonian Formalism}

It is clear from our previous section that, for the absolute anticommutativity of the co-BRST and anti-co-BRST symmetry transformations, one has to invoke a CF type restriction (i.e ${\cal{B}}_\mu - \bar{{\cal{B}}}_\mu -\partial_\mu\varphi _2=0$). For this condition, to persist with respect to the time-evolution of our gauge system, it is essential requirement that it should remain time invariant quantity. To this goal in mind, it can be seen that the following canonical Hamiltonian densities emerge from the Lagrangian densities (\ref{l4}) and 
(\ref{l5}):
\begin{eqnarray}
&&{\cal{H}}^{(4)} =
(\Pi _{ij}^{(4)})^2+ 2\;({\Pi _{\varphi _1}^{(4)}})^2 - 2\;({\Pi _{\varphi _2}^{(4)}})^2 -\frac{1}{2}\;({\Pi _{0i}^{(4)}})^2 + 2\;\Pi _{ij}^{(4)}\;
\partial _i B_{0j}+\frac{1}{2}\;(\Pi _{0i}^{(4)})\;\partial _i \varphi _1 \nonumber\\
&&-(\Pi_{0j}^{(4)})\; \partial
 _i B_{ij} + 2 \;(\Pi _{\varphi _1}^{(4)})\; (\partial _i B_{i0}) + ({\Pi _{\varphi _2}^{(4)}})\epsilon _{ijk}\;\partial_i B_{jk} +\frac{1}{2}\;\epsilon _{ijk}\;\Pi_{jk}^{(4)}\;\partial_i \varphi _2
+ {\cal H}_{(g)}, \label{h4}
\end{eqnarray}
\begin{eqnarray}
&& {\cal{H}}^{(5)} =
(\Pi _{ij}^{(5)})^2+ 2\;({\Pi _{\varphi _1}^{(5)}})^2 - 2\;({\Pi _{\varphi _2}^{(5)}})^2 -\frac{1}{2}\;({\Pi _{0i}^{(5)}})^2 + 2\;\Pi _{ij}^{(5)}\;
\partial _i B_{0j}+\frac{1}{2}\;(\Pi _{0i}^{(5)})\;\partial _i \varphi _1 \nonumber\\
&&-(\Pi_{0j}^{(5)})\; \partial
 _i B_{ij} - 2 \;(\Pi _{\varphi _1}^{(5)})\; (\partial _i B_{i0})- ({\Pi _{\varphi _2}^{(5)}})\epsilon _{ijk}\;\partial_i B_{jk} +\frac{1}{2}\;\epsilon _{ijk}\;\Pi_{jk}^{(5)}\;\partial_i \varphi _2
+ {\cal H}_{(g)}, \label{h5}
\end{eqnarray}
where the canonical momenta are defined as:
\begin{eqnarray}
\Pi_{\varphi _1}^{(4)} &\equiv & \frac{\partial{{\cal{L}}^{(4)}}}{\partial (\partial_0{\varphi _1})} 
= \frac{B_0}{2}, \;\quad \quad \quad \quad \quad \Pi_{\varphi _1}^{(5)}\equiv \frac{\partial{\cal{L}}^{(5)}}{
\partial(\partial_0\varphi _1)} = -\frac{\bar{B}_0}{2}, \nonumber\\
\Pi_{\varphi _2}^{(4)}&\equiv & \frac{\partial{\cal{L}}^{(4)}}{\partial(\partial_0\varphi _2)} = - 
\frac{{\cal{B}}_0}{2}, \;\quad \quad \quad \quad \; \Pi_{\varphi _2}^{(5)}\equiv \frac{\partial{\cal{L}}^{(5)}}{
\partial(\partial_0\varphi _2)} = \frac{\bar{{\cal{B}}}_0}{2}, \nonumber\\
\Pi_{0i}^{(4)}&\equiv & \frac{\partial{{\cal{L}}^{(4)}}}{\partial (\partial_0{B _{0i})}}= B_i,
\quad \quad \quad \quad \quad \Pi_{0i}^{(5)}\equiv \frac{\partial{{\cal{L}}^{(5)}}}{\partial (\partial_0{B _{0i})}} =  \bar{B}_i, \nonumber\\
\Pi_{ij}^{(4)}&\equiv & \frac{\partial{{\cal{L}}^{(4)}}}{\partial (\partial_0{B _{ij})}}= -\frac
{1}{2} \epsilon _{ijk}{\cal{B}}_k,\;\quad\quad \Pi_{ij}^{(5)}\equiv \frac{\partial{{\cal{L}}^{(5)}}}{\partial (\partial_0{B _{ij})}}= - \frac
{1}{2} \epsilon _{ijk}\bar{{\cal{B}}}_k.
\end{eqnarray}
It will be noted that the superscripts (``$(4)$ and $(5)"$ ) on the Hamiltonian densities and momenta correspond to such superscripts on the Lagrangian densities (\ref{l4}) and (\ref{l5}).
The equations of motion, derived from the Heisenberg's equation of motion 
(with $ H^{(4,5)} = \int d^3 x \;{\cal{H}}^{(4,5)})$, are found to be exactly same as the following juxtaposed Euler-Lagrange equation of motion derived from the Lagrangian densities (\ref{l4}) and (\ref{l5}), namely
\begin{eqnarray}
&B_\mu =\partial^\nu B_{\nu\mu} +\frac{1}{2}\partial_\mu \varphi _1,\; \quad \quad \;\; \bar{B}_\mu= 
\partial^\nu B_{\nu\mu} -\frac{1}{2}\partial_\mu \varphi _1, &\nonumber\\
&{\cal{B}}_\mu =\frac{1}{2}\varepsilon _{\mu \nu \eta \kappa }\partial^\nu B^{\eta \kappa }+ 
\frac{1}{2}\partial_\mu \varphi _2,\;\quad \quad\;\; \bar{{\cal{B}}}_\mu =\frac{1}{2}\varepsilon 
_{\mu \nu \eta \kappa }\partial^\nu B^{\eta \kappa }- \frac{1}{2}\partial_\mu \varphi _2, &
\nonumber\\
&\partial_\mu {\cal{B}}_\nu - \partial_\nu {\cal{B}}_\mu - \varepsilon _{\mu \nu \eta \kappa }
\partial^\eta B^\kappa= 0, \quad \quad \partial_\mu \bar{{\cal{B}}}_\nu - \partial_\nu\bar{{
\cal{B}}}_\mu - \varepsilon _{\mu \nu \eta \kappa }\partial^\eta \bar{B}^\kappa= 0, &\nonumber\\
&\partial_\mu B_\nu - \partial_\nu B_\mu - \varepsilon _{\mu \nu \eta \kappa }\partial^\eta 
{\cal{B}}^\kappa= 0,\; \quad \quad \partial_\mu \bar{B}_\nu - \partial_\nu \bar{B}_\mu - 
\varepsilon _{\mu \nu \eta \kappa }\partial^\eta \bar{{\cal{B}}}^\kappa= 0, &\label{el1}
\end{eqnarray}
where the left set of equations are from (\ref{l4}) and that of the right are from (\ref{l5}). Exactly the above set of equations can be derived from the Hamiltonian densities (\ref{h4}) and (\ref{h5}) which are explicitly given in our Appendix C.

It is worthwhile to mention that the CF type restrictions (${\cal{B}}_\mu - 
\bar{{\cal{B}}}_\mu -\partial_\mu \varphi _2=0,\quad {\cal{B}}_\mu + \bar{{\cal{B}}}_\mu 
-\varepsilon _{\mu \nu \eta \kappa }\partial^\nu B^{\eta \kappa}=0$) invoked for the proof 
of the absolute anticommutativity of the (anti-) dual-BRST symmetry transformations, are 
derived in two steps and they cannot emerge from a single Lagrangian and/or Hamiltonian 
densities. We achieve this goal below and show that a single Lagrangian density (and the 
corresponding Hamiltonian density) can produce the CF type restrictions in one step.

Besides the re-definitions in  (\ref{af1}), we re-define the following auxiliary fields
\begin{equation}
h_\mu = \frac{1}{2} ({\cal{B}}_\mu +\bar{{\cal{B}}}_\mu), \;
\quad \quad \bar{h}_\mu = \frac{1}{2} ({\cal{B}}_\mu - \bar{{\cal{B}}}_\mu),
\end{equation}
to express the following Lagrangian density (cf. (\ref{l4}) and (\ref{l5})) as:
\begin{eqnarray}
{\cal{L}}^{(6)}&=& \frac{1}{2} \;\left( {\cal{L}}^{(4)}+ {\cal{L}}^{(5)}\right)
\quad \equiv \quad \frac{1}{2}\;(h\cdot h+\bar{h}\cdot\bar{h}) - \frac{1}{2}\;h^\mu \;\varepsilon_ {\mu \nu \eta 
\kappa }\;\partial^\nu B^{\eta\kappa}\nonumber\\
 & - &\frac{1}{2}\;\bar{h}_\mu\;\partial_\mu\varphi _2
+ b^\mu\;(\partial^\nu B_{\nu\mu}) + \frac{1}{2}\;\bar{b}^\mu\; \partial_\mu \varphi _1 - \frac
{1}{2}\;(b\cdot{b}+\bar{b}\cdot{\bar{b}}) + {\cal{L}}_{(g)}, \label{l6}
\end{eqnarray}
where we have used
\begin{equation}
({\cal{B}}\cdot{\cal{B}} + \bar{{\cal{B}}}\cdot\bar{{\cal{B}}})= 2\;(h\cdot h +\bar{h}\cdot \bar
{h}).
\end{equation}
The following Euler-Lagrange equations of motion emerge from 
(\ref {l6}):
\begin{eqnarray}
&\bar{b}_\mu =\frac{1}{2} \partial_\mu\varphi _1,\;
 \qquad \bar{h}_\mu = \frac{1}{2}\partial_\mu\varphi _2,
\; \qquad b_\mu=\partial^\nu B_{\nu\mu}, \qquad h_\mu
= \frac{1}{2} \varepsilon _{\mu \nu \eta \kappa }\partial^\nu B^{\eta\kappa}, &\nonumber\\
&\partial\cdot\bar{b}= 0, \;\qquad \partial\cdot\bar{h}=0, \qquad \Box\varphi _1 = 0,\;\qquad \Box\varphi _2=0, \;
\qquad \partial \cdot b = 0,\;\qquad \partial\cdot h = 0, &\nonumber\\
&\varepsilon _{\mu \nu \eta \kappa }\partial^\eta h^\kappa + (\partial^\eta b^\kappa-\partial^\kappa b^\eta)= 0, &
\end{eqnarray}
besides the ghost field equations that are derived from ${\cal{L}}_{(g)}$.
The canonical momenta, from (\ref{l6}), are:
\begin{equation}
\Pi_{\varphi _1}= \frac{\bar{b}_0}{2},\;\qquad \;\Pi_{\varphi _2}= -\frac{\bar{h}_0}{2},\; \qquad \;\Pi_{0i}= b_i,\; \quad \Pi_{ij}=-\frac{1}{2}\epsilon _{ijk}\;h_k.
\end{equation}
It is  evident that $\Pi_\mu^{(b)}=0,\Pi_\mu^{(h)}=0,\Pi_\mu^{(\bar{h})}=0, \Pi_\mu^{(\bar{b})}=0,$ because $b_\mu, \bar{b}_\mu,h_\mu,\bar{h}_\mu$ are the auxiliary fields of the theory.

At this juncture, it can be seen that $\bar{h}_\mu =\frac{1}{2}\partial_\mu\varphi _2,$ and $h_\mu =\frac{1}{2} \varepsilon _{\mu \nu \eta \kappa }\partial^\nu B^{\eta\kappa}$ lead to the CF type of restrictions: ${\cal{B}}_\mu -\bar{\cal{B}}_\mu-\partial_\mu\varphi _2=0$ and ${\cal{B}}_\mu +\bar{\cal{B}}_\mu-\varepsilon _{\mu \nu \eta \kappa }\partial^\nu B^{\eta\kappa}=0$ in a single step and they are derived from a single Lagrangian density (i.e ${\cal{L}}^{(6)}$) that is obtained from the linear combination of ${\cal{L}}^{(4)}$ and ${\cal{L}}^{(5)}$. It will be noted that the other linear combination [${\cal{L}}^{(4)}- {\cal{L}}^{(5)}$] does not lead to an interesting Lagrangian density because the ghost parts of the Lagrangian densities $ {\cal{L}}^{(4,5)}$ cancel out with each other in this combination.

The canonical Hamiltonian density, emerging from the Lagrangian density ${\cal{L}}^{(6)}$, is  
\begin{eqnarray}
{\cal{H}}^{(6)} &=& \Pi_{ij}^2 -2\; \Pi_{\varphi _2}^2 -\frac{1}{2}\;(\Pi_{0i})^2 +2\;\Pi_{
\varphi _1}^2 +\frac{1}{2}\;(b_0 b_0-\bar{b}_i \bar{b}_i) -\frac{1}{2}\;(h_0 h_0 - \bar{h}_i \bar
{h}_i)+\frac{1}{2}\;\bar{b}_i\;\partial_i\varphi _1 \nonumber\\
&-& \frac{1}{2}\;\bar{h}_i\;\partial_i\varphi _2
 -  b_0\;\partial_i B_{0i} - \Pi_{0j}\;\partial_i B_{ij} + 2\;\Pi_{jk}\;\partial_j B_{0k}-
\frac{h_0}{2}\;
\epsilon _{ijk}\;\partial_i B_{jk} +  {\cal H}_{(g)}. \label{hb6}
\end{eqnarray}
It will be noted that, corresponding to the auxiliary fields $b_0,h_o,\bar{b}_i,\bar{h}_i,$ there are no momenta in the above expression because these are the primary constraints on the theory (i.e $\Pi_0^{(b_0)}\approx 0, \Pi_i^{(\bar{b})}\approx 0,\Pi_0^{h_0}\approx0,\Pi_i^{(\bar{h})}\approx 0, $ ). It is straightforward to check that the time evolution invariance of these constraints (with $H^{(6)}= \int d^3x {\cal{H}}^{(6)}$):
\begin{eqnarray}
{\dot{\Pi}}_0^{(b_0)}&=& -i \left[\Pi_0^{(b_0)},H^{(6)}\right ] \;=0 \quad \Rightarrow \quad b_0=\partial^i 
B_{i0}, \nonumber\\
{\dot{\Pi}}_i^{(\bar{b})}&=& -i \left[\Pi_i^{(\bar{b})},H^{(6)}\right ] \;=0 \;\quad \Rightarrow
\quad \bar{b}_i = \frac{1}{2}\partial_i \varphi _1, \nonumber\\
{\dot{\Pi}}_0^{(h_0)}&=& -i \left[\Pi_0^{(h_0)},H^{(6)}\right ] =0 \;\quad \Rightarrow \quad h_0= -
\frac{1}{2}\epsilon _{ijk}\partial_i B_{jk}, \nonumber\\
{\dot{\Pi}}_i^{(\bar{h})}&=& -i \left[\Pi_i^{(\bar{h})},H^{(6)}\right ] \;\;=0 \quad \Rightarrow \quad \bar{h}_i=\frac{1}{2}\partial_i \varphi _2, \label{cf2}
\end{eqnarray}
leads to the CF type restrictions $B_0+\bar{B}_0 - 2\partial^i B_{i0}=0,\; B_i-\bar{B}_i-\partial_i\varphi _1=0,\;{\cal{B}}_0+\bar{{\cal{B}}}_0+\epsilon _{ijk}\partial_i B_{jk}=0,\; {\cal{B}}_i+\bar{{\cal{B}}}_i-\partial_i\varphi _2=0$ which are like the secondary constraints on the theory.

The full set of CF type restrictions (i.e $B_\mu -\bar{B}_\mu-\partial_\mu\varphi _1=0,\; B_\mu+\bar{B}_\mu-2\partial^\nu B_{\nu\mu}=0,\; {\cal{B}}_\mu-\bar{{\cal{B}}}_\mu-\partial_\mu\varphi _2=0,\;{\cal{B}}_\mu+\bar{{\cal{B}}}_\mu-\varepsilon _{\mu \nu \eta \kappa }\;\partial^\nu B^{\eta\kappa}=0 $) can be obtained from the Hamiltonian ($H^{(6)}$) if we invoke the time- evolution of the following basic fields:
\begin{eqnarray}
\dot{\varphi _1}&=& -i\left[\varphi _1,H^{(6)}\right ]\quad \; \Rightarrow \quad \bar{b}_i=\frac{1}{2}\partial_0\varphi _1, \nonumber\\
\dot{\varphi _2}&=& -i\left[\varphi _2,H^{(6)}\right ]\quad \; \Rightarrow \quad \bar{h}_0=\frac{1}{2}\partial_0\varphi _2, \nonumber\\
\dot{B}_{0i}&=& -i\left[B_{0i},H^{(6)}\right ]\quad \Rightarrow \quad b_i=\partial^\mu B_{\mu i}, \nonumber\\
\dot{B}_{ij}&=& -i\left[B_{ij},H^{(6)}\right ]\quad \Rightarrow \quad h_i=-\frac{1}{2}\epsilon _{ijk}\partial_0 B_{jk}-\epsilon _{ijk}\partial_j B_{k0}, \label{cf1}
\end{eqnarray}
in addition to the expression obtained in (\ref {cf2}).
Thus, we note that it is the combination of (\ref{cf1}) and (\ref{cf2}) that yields all the components of the CF type restriction that are invoked in the proof of the absolute anticommuatativity of the nilpotent symmetry transformations.

In its full glory, the total Hamiltonian density is the sum of the canonical Hamiltonian density (\ref{hb6}) and the primary constraints on the theory as given below.
\begin{equation}
{\cal{H}}^{(6)}_T =\Pi_0^{(b_0)}\partial_0 b_0 +\Pi_0^{(h_0)}\partial_0 h_0 -\Pi_i^{(\bar{b})}\partial_0\bar{b}_i-\Pi_i^{(\bar{h})}\partial_0\bar{h}_i +{\cal{H}}^{(6)}.
\end{equation}
Time-evolution invariance of the CF type restrictions ( ${\cal{B}}_\mu-\bar{{\cal{B}}}_\mu-\partial_\mu\varphi _2=0 ,{\cal{B}}_\mu+\bar{{\cal{B}}}_\mu-\varepsilon _{\mu \nu \eta \kappa }\partial^\nu B^{\eta\kappa}=0 $) can be now checked to be true with the total Hamiltonian density ${\cal{H}}_T^{(6)}$. Infact, using the canonical brackets, it is quite straightforward to check that 
\begin{eqnarray}
&\left[{\cal{B}}_0-\bar{{\cal{B}}}_0-\partial_0 \varphi _2, {\cal{H}}_T^{(6)}\right ]= 0,\;\quad \quad 
\left[{\cal{B}}_i-\bar{{\cal{B}}}_i - \partial_i \varphi _2,{\cal{H}}_T^{(6)}\right ]=0, &\nonumber\\
&\left[{\cal{B}}_0+\bar{{\cal{B}}}_0-\epsilon _{ijk}\partial_i B_{jk},{\cal{H}}_T^{(6)}\right ]=0,\;
\quad \quad \left[{\cal{B}}_i+\bar{{\cal{B}}}_i + \epsilon _{ijk}(\partial_0 B_{jk} 
+ 2\partial_j B_{k0}),{\cal{H}}_T^{(6)}\right ] = 0.&
\end{eqnarray}
The above relations show that the CF type restrictions remain the same during the full time-evolution of the 2-form Abelian gauge system. As a consequence, it is proper to impose these conditions for the proof of the absolute anticommutativity of the dual-BRST and anti-dual BRST symmetries during the full dynamical evolution
of our present free Abelian 2-form gauge theory in physical four dimensions of spacetime.

\section{Conclusions}

In our present investigation, we have concentrated on the dynamical aspects of the 4D free Abelian 2-form gauge theory in the framework of the Hamiltonian formulation. This field theoretic model happens to be the 
off-shell nilpotent (anti-) BRST as well as (anti-) co-BRST invariant model of a 4D gauge theory. We have 
derived the dynamical equations of the theory with the help of the Heisenberg equations of motion where 
the Hamiltonian (of the (anti-) BRST as well as (anti-)co-BRST invariant system) plays a central role.

Our earlier works [9-11,21-23], devoted to the discussion of the Abelian 2-form gauge theory, have been carried out in the Lagrangian formulation where the CF type restrictions have been derived as the Euler-Lagrange equations of motion from the coupled Lagrangian densities. These CF type restrictions are required for the proof of an absolute anticommutativity between the off-shell nilpotent

     (i) BRST and anti-BRST symmetry transformations, and
     
     (ii) co-BRST and anti-co-BRST smmetry transformations.\\
However, the Lagrangian formulation does not shed any light on the time-evolution invariance of the above CF type restrictions.

We have chosen, in our present endeavour, the Hamiltonian formalism so that we can clearly demonstrate that the CF type restrictions remain invariant w.r.t time-evolution of the Abelian 2-form gauge system. This result provides a logical reason behind the imposition of the CF type restrictions which are valid at any moment of time for the full time-evolution of our physical 2-form Abelian gauge system in 4D spacetime.

The key difference between our present endeavour and our earlier attempts \cite{guma,hama} is the fact that CF type restrictions, that are at the heart of the absolute anticommutativity of the (anti-) BRST and (anti-) co-BRST symmetry transformations, are derived from a single Lagrangian density (and corresponding Hamiltonian density) in a single step. This should be contrasted with our earlier Lagrangian formulation where a set of coupled Lagrangian densities led to the derivations of the CF type restrictions in two steps as the Euler-Lagrange equations of motion and their subtraction/addition.

The absolute anticommutativity of the nilpotent (anti-) BRST and (anti-) co-BRST symmetry transformations is an essential requirement because it ensures the linear independence of the (i) BRST versus anti-BRST and (ii) co-BRST versus anti-co-BRST symmetries. Furthermore, it confirms physically
the independent roles of the anti-BRST symmetries and anti-co-BRST symmetries in the context of the 4D Abelian 2-form gauge theory. It will be recalled that the anti-BRST and anti-co-BRST symmetries do not play any independent role 
{\it vis-{\`a}-vis} the BRST and co-BRST symmetries in the context of the 4D Abelian 1-form\footnote{In the case of 4D Abelian 1-form gauge theory, the operator form of the first-class constraints annihilate the physical
states of the theory due to the physicality criteria ($Q_{(a)b} |phys> = 0$) with the (anti-) BRST charges $Q_{(a)b}$. In other words, the BRST and anti-BRST charges lead to the same conditions 
through $Q_{(a)b} |phys> = 0$. Thus, the anti-BRST charge does not play an independent role here.} 
gauge theory (see, e.g. [24]). These points are consistent with the results of our work on superfield formulation of the Abelian 2-form gauge theory \cite{rpm3}.

One of us has studied the gauge theories in BRST superspace [25-29], which is slightly different from the usual approach of the superspace formulation (see, e.g. [15]). The main features of 
this BRST superspace are (i) the whole action, including the source terms for the composite operators, is accommodated in a single compact superspace action, (ii) theory has generalized gauge invariance and WT identities which are realised in a simple way, and (iii) operation like super-rotation and super-translation, in anticommuting variable, can be carried out in a completely unrestricted manner. Such superspace formulation is very useful in studying the renormalization problem in gauge theories. It would be nice endeavour to apply this approach to study the 2-form \cite{dema} and higher-form gauge theories.

To generalize our present work and earlier works [9-11,21-23] to 4D {\it non-Abelian} 2-form and higher p-form (p$>2$) gauge theories is one of the challenging future endeavour. We expect that even the higher-form (p$>2$) {\it Abelian} gauge theories would lead to some very interesting observations in the framework of BRST formalism. A
thorough constraint analysis of our current theory\footnote{Only a few comments have been
made by us on the constraints of our present theory. However, an elaborate discussion on the classification 
of these constraints and their specific roles, in the context of our present theory, would be taken up
in our future endeavour [30].}
and higher-form gauge theories is also on our future agenda.
 Discussion of the above theories in the framework of superfield formulations [15,25-29] is yet another direction for future investigation. Currently these problems are under investigation and our results would be reported in our future publications [30].\\

\noindent
{\bf Acknowledgements:} \\

\noindent
One of us (RPM) thankfully acknowledges the financial support
from the Department of Science and Technology (DST), Government of India, under
the SERC project sanction grant No. SR/S2/HEP-23/2006.\\

\begin{center}
{\large{\bf{Appendix A}}}
\end{center}

We explicitly demonstrate that the Hamiltonian $H_{(g)} = \int d^3x \; {\cal{H}}_{(g)}$, 
corresponding to the ghost part ${\cal{L}}_{(g)} $ (cf. (\ref{lg})) of the Lagarangian densities (\ref{l1}) and (\ref{l2}), 
yields all the equations of motion (cf. (\ref{ag})) for the (anti-) ghost fields. For this purpose,
we have to exploit the 
following canonical (anti-) commutators (with $\hbar = c = 1$):
\begin{eqnarray}
\left[\beta({\bf{x}}, t) , \Pi^{(\beta)}({\bf{y}}, t)\right ]&=& i\delta ^{(3)}({\bf{x}}-{\bf
{y}}),\nonumber\\
\left[\bar{\beta}({\bf{x}},t) , \Pi^{(\bar{\beta})}({\bf{y}},t)\right ]&=& i\delta ^{(3)}({\bf
{x}}-{\bf{y}}), \nonumber\\
\left\{C_0({\bf{x}},t) ,  \Pi_0^{(c_0)}({\bf{y}},t)\right\}&=& i\delta ^{(3)}({\bf
{x}}-{\bf{y}}), \nonumber\\
\left\{\bar{C}_0({\bf{x}},t) ,  \Pi_0^{(\bar{c}_0)}({\bf{y}},t)\right\}&=& i\delta ^{(3)}({\bf
{x}}-{\bf{y}}), \nonumber\\
\left\{C_i({\bf{x}},t) ,  \Pi^{(c)}_j({\bf{y}},t)\right\}&=& - i \delta _{ij}\delta ^{(3)}({\bf
{x}}-{\bf{y}}), \nonumber\\
\left\{\bar{C}_i({\bf{x}},t) ,  \Pi^{(\bar{c})}_j({\bf{y}},t)\right\}&=& - i \delta _{ij}  
\delta ^{(3)}({\bf{x}}-{\bf{y}}), \label{ac}
\end{eqnarray}
and all the other (anti-) commutators are zero.

Using (\ref{ac}), it can be checked that the time-evolution of the canonical momenta 
\begin{eqnarray}
{\dot{\Pi}}^{(\beta)} &=& - i \left[\Pi ^{(\beta)}, H_{(g)}\right ] \quad \Rightarrow \quad \Box
{\bar{\beta}} = 0,  \nonumber\\
{\dot{\Pi}}^{(\bar{\beta})} &=& - i \left[\Pi ^{(\bar{\beta})}, H_{(g)}\right ] \quad 
\Rightarrow \quad \Box{\beta} =0,  \nonumber\\
{\dot{\Pi}}^{(c_0)}_0 &=& \; + i \left[\Pi ^{(c_0)}_0 , H_{(g)}\right ] \quad \Rightarrow \quad \Box{
\bar{C}_0} = -\partial _0 \rho,   \nonumber\\
{\dot{\Pi}}^{(\bar{c}_0)}_0 &=& \;+  i\left[\Pi ^{(\bar{c}_0)}_0 , H_{(g)}\right ] \quad \Rightarrow
\quad \Box{C_0} = \partial _0 \lambda,   \nonumber\\
{\dot{\Pi}}^{(c)}_i &=& \; +  i\left[\Pi ^{(c)}_i ,  H_{(g)}\right ] \;\quad \Rightarrow \quad \Box{\bar{
C}_i} = -\partial _i \rho,  \nonumber\\
{\dot{\Pi}}^{(\bar{c})}_i &=& \; \; + i\left[\Pi ^{(\bar{c})}_i ,  H_{(g)}\right ] \;\quad \Rightarrow 
\quad \Box{C_i} = \partial _i \lambda,  
\end{eqnarray} 
lead to the Euler-Lagrange equation of motion derived from the Lagrangian densities (\ref{l1}) and/or (\ref{l2}) for the basic (fermionic) bosonic (anti-) ghost fields of the theory.

On the other hand, it is interesting that the time-evolution of the (anti-) ghost fields 
\begin{eqnarray}
\dot{\beta} &=& -i \left[ \beta , H_{(g)}\right] \;\quad \Rightarrow \quad \dot{\beta}=\Pi^{(\bar{\beta})},\nonumber\\
\dot{\bar{\beta}} &=& -i \left[ \bar{\beta} , H_{(g)}\right] \;\quad \Rightarrow \quad \dot{\bar{
\beta}}=\Pi^{(\beta)}, \nonumber\\
\dot{C}_0 &=& \; + i \left[ C_0 , H_{(g)}\right] \quad \Rightarrow \quad \Pi^{(\bar{c}_0)}_0=\frac{1}{2}(\partial \cdot C)=\lambda, \nonumber\\
\dot{\bar{C}}_0 &=& \;+ i \left[ \bar{C}_0 , H_{(g)}\right] \quad \Rightarrow \quad \Pi^
{(c_0)}_0= -\frac{1}{2}(\partial \cdot \bar{C})=\rho,  \nonumber\\
\dot{C}_i &=& + i \;\;\left[ C_i , H_{(g)}\right] \quad \Rightarrow \quad \Pi^{(\bar{c})}_i=(\partial_0 C_i - \partial_i C_0 ), \nonumber\\
\dot{\bar{C}}_i &=& \;\;+  i \left[ \bar{C}_i , H_{(g)}\right] \quad \Rightarrow \quad \Pi^{(c)}_i= -(\partial_0 \bar{C}_i - \partial_i \bar{C}_0 ), 
\end{eqnarray}
leads to the definition of the canonical momenta corresponding to the bosonic and fermionic (anti-) ghost 
fields. This establishes the consistency and equivalence between the Lagrangian and 
Hamiltonian descriptions of the Abelian 2-form gauge theory.\\
\\
\begin{center}
{\large{\bf{Appendix B}}}
\end{center}
Dynamics of the non-ghost part of the Lagrangian densities (\ref{l1}) and (\ref{l2}) remain unaffected due to their description in the framework of Lagrangian and Hamiltonian formalism. To establish this fact, it can be checked that the Hamiltonian $H_{(b)}^{(1)} = \int d^3x\; {\cal{H}}_{(b)}^{1)}$, produces the following Hesisenberg dynamical equations of motion for the basic fields:
\begin{eqnarray}
\dot{\varphi}_1  & = & \; -i \left[\varphi _1 , H_{(b)}^{(1)}\right ]\quad \Rightarrow \quad B_0 = 
\frac{1}{2}\partial _0 \varphi _1 +\partial _i B_{0i}, \nonumber\\
\dot{B_{oi}} &=& -i\left [B_{0i} ,H_{(b)}^{(1)}\right ] \quad \Rightarrow \quad B_i =\partial^\nu B_{
\nu i}  +\frac{1}{2} \partial_i \varphi_1, \nonumber\\
\dot{B}_{ij} &=& -i\left[ B_{ij} ,H_{(b)}^{(1)}\right ] \quad \Rightarrow \quad \Pi_{ij}=\frac{1}{2} 
H_{0ij}, \label{e1}
\end{eqnarray}
where we have exploited the canonical brackets (\ref{cb}).

On the other hand, the time evolution of the canonical momenta, namely;
\begin{eqnarray}
\dot{\Pi}_{\varphi_1}^{(1)} & = & -i\left [\Pi_{\varphi _1}^{(1)} , H_{(b)}^{(1)}\right ] \quad
\Rightarrow \quad \partial \cdot  B = 0, \nonumber\\
\dot{\Pi}_{0i}^{(1)} &=& -i\left [ \Pi_{0i}^{(1)} , H_{(b)}^{(1)}\right ] \quad \Rightarrow \quad
\partial_k H^{k0i} + \partial^0 B^i -\partial^i B^0=0, \nonumber\\
\dot{\Pi}_{ij} &=& -i\left[\Pi_{ij} ,H_{(b)}^{(1)}\right ] \;\quad \Rightarrow \quad\partial_\mu H^{\mu ij} + (\partial^i B^j -\partial^j B^i)=0, \label{e2}
\end{eqnarray}
produces the dynamical equations of motion. It will be noted that the top two equations of (
\ref{e1}) and bottom two equations of (\ref{e2}) can be combined together as : $ B_\mu = 
\partial^\nu B_{\nu\mu}+\frac{1}{2}\partial_\mu\varphi _1 $, $ \partial_\mu H^{\mu \nu 
\kappa } +\partial^\nu B^\kappa - \partial^\kappa B^\nu =0 $. These finally lead to the simple equations of motion: 
$\Box{\varphi _1} =0$  (due to $\partial\cdot B=0$) and $\Box{B_{\mu \nu }}=0 $ as well as 
$\Box{B_\mu}=0$.

Similarly, the Hamiltonian $H_{(b)}^{(2)} = \int d^3x \;{\cal{H}}_{(b)}^{(2)}$ leads to the following equations of motion (that are different from $H_{(b)}^{(1)}$ ), namely;
 \begin{eqnarray}
\dot{\varphi}_1  & = &  -i \left[\varphi _1 , H_{(b)}^{(2)}\right ]\;\;\quad \Rightarrow \quad \bar
{B}_0 =  \partial _i B_{i0} - \frac{1}{2}\partial _0 \varphi_1,  \nonumber\\
\dot{B_{oi}} &=& -i\left [B_{0i} ,H_{(b)}^{(2)}\right ] 
\;\quad \Rightarrow \quad \bar{B}_i =
\partial^\nu B_{\nu i}  - \frac{1}{2} \partial_i \varphi_1, \nonumber\\
\dot{\Pi}_{0i}^{(2)} &=& -i\left [ \Pi_{0i}^{(1)} , H_{(b)}^{(2)}\right ] \quad \Rightarrow 
\quad \partial_k H^{k0i} + \partial^0 \bar{B}^i -\partial^i \bar{B}^0 = 0, \nonumber\\
\dot{\Pi}_{ij} &=& -i\left[\Pi_{ij} ,H_{(b)}^{(2)}\right ] \;\quad \Rightarrow \quad \partial_\mu H^{
\mu ij} + (\partial^i \bar{B}^j -\partial^j \bar{B}^i)=0, \nonumber\\
\dot{\Pi}_{\varphi_1}^{(2)} & = & -i\left [\Pi_{\varphi _1}^{(2)} , H_{(b)}^{(2)}\right ] \quad
\Rightarrow \quad \partial \cdot \bar{B}=0. 
\end{eqnarray}
Ultimately, the above equation imply that $\Box{\varphi _1}=0 , \Box{B_{\mu\nu}}=0,$ and $ \Box{\bar{B_\mu}}=0$. These equations primarily emerge from $\bar{B}_\mu = \partial^\nu B_{\nu\mu}-\frac{1}{2}\partial_\mu\varphi _1$ and $\partial_\mu H^{\mu \nu \kappa } 
+\partial^\nu \bar{B}^\kappa - \partial^\kappa\bar{B}^\nu=0 $.

\begin{center}
{\bf  Appendix C}
\end{center}

The Euler-Lagrange equations of motion (\ref{el1}) can be re-derived from the Hamiltonians $ H^{(4,5)}=\int d^3x\; {\cal{H}}^{(4,5)}$ as illustrated below 
\begin{eqnarray}
\dot{\varphi}_1  & = &  -i \left[\varphi _1 , H^{(4)}\right ]\;\;\quad \Rightarrow \quad B_0 = \frac{1}{2}
\partial _0 \varphi _1 -\partial _i B_{i0}, \nonumber\\
\dot{\varphi}_2 & = & -i \left[\varphi _2 , H^{(4)}\right ] \;\;\quad \Rightarrow \quad {\cal{B}}_0 =\frac
{1}{2}\partial _0 \varphi _2 - \frac{1}{2} \epsilon _{ijk}\partial _i B_{jk}, \nonumber \\
\dot{\Pi}_{\varphi_1}^{(4)} & = & -i\left [\Pi_{\varphi _1}^{(4)} , H^{(4)}\right ]\quad \Rightarrow \quad
\partial \cdot B =0,  \nonumber\\
\dot{\Pi}_{\varphi_2}^{(4)} & = & -i\left [\Pi_{\varphi _2}^{(4)} , H^{(4)}\right ]\quad \Rightarrow \quad
\partial \cdot{\cal{B}}=0, \nonumber\\
\dot{B_{oi}} &=& \; -i\left [B_{0i} ,H^{(4)}\right ] \quad \Rightarrow \quad B_i =\partial_0 B_{
0i}-\partial_k B_{ki}  +\frac{1}{2} \partial_i \varphi_1, \nonumber\\
\dot{B}_{ij} &=& \; -i\left[ B_{ij} ,H^{(4)}\right ] \quad \Rightarrow \quad {\cal{B}}_i = \frac{1}{2} 
\epsilon _{ijk}(\partial_j B_{0k}- \partial_k B_{0j} - \partial_0 B_{jk}) +\frac{1}{2} \partial_i \varphi _2, \nonumber\\
\dot{\Pi}_{ij}^{(4)} &=& -i\left[\Pi_{ij}^{(4)} ,H^{(4)}\right ]\quad \Rightarrow \quad \partial_0 {
\cal{B}}_i -  \partial_i{\cal{B}}_0 - \epsilon _{ijk}\partial_j B_k=0, \nonumber\\
\dot{\Pi}_{0i}^{(4)} &=& -i\left[\Pi_{0i}^{(4)} ,H^{(4)}\right ] \quad \Rightarrow \quad \partial_0 B_i 
-  \partial_i B_0 + \epsilon _{ijk}\partial_j {\cal{B}}_k =0,
\end{eqnarray} 
and the Hamiltonians $H^{(5)}=\int d^3x \;{\cal{H}}^{(5)} $ leads to:
\begin{eqnarray}
\dot{\varphi}_1  & = &  -i \left[\varphi _1 , H^{(5)}\right ] \;\;\quad \Rightarrow \quad \bar{B}_0 = \partial _i B_{i0}-\frac{1}{2}
\partial _0 \varphi _1,  \nonumber\\
\dot{\varphi}_2 & = & -i \left[\varphi _2 , H^{(5)}\right ] \;\;\quad \Rightarrow \quad \bar{{\cal{B}}}_0 =
-\frac
{1}{2}\partial _0 \varphi _2 - \frac{1}{2} \epsilon _{ijk}\partial _i B_{jk}, \nonumber \\
\dot{\Pi}_{\varphi_1}^{(5)} & = & -i\left [\Pi_{\varphi _1}^{(5)} , H^{(5)}\right ]\quad \Rightarrow \quad
\partial \cdot \bar{B} =0, \nonumber\\
\dot{\Pi}_{\varphi_2}^{(5)} & = & -i\left [\Pi_{\varphi _2}^{(5)} , H^{(5)}\right ]\quad \Rightarrow \quad
\partial \cdot\bar{{\cal{B}}}=0, \nonumber\\
\dot{B_{oi}} &=& \;-i\left [B_{0i} ,H^{(5)}\right ] \quad \Rightarrow \quad \bar{B}_i =\partial_0 B_{
0i}-\partial_k B_{ki}  -\frac{1}{2} \partial_i \varphi_1, \nonumber\\
\dot{B}_{ij} &=& \;-i\left[ B_{ij} ,H^{(5)}\right ] \quad \Rightarrow \quad \bar{{\cal{B}}}_i = \frac{1}{2} 
\epsilon _{ijk}(\partial_j B_{0k}- \partial_k B_{0j} - \partial_0 B_{jk}) -\frac{1}{2} \partial_i \varphi _2, \nonumber\\
\dot{\Pi}_{ij}^{(5)} &=& -i\left[\Pi_{ij}^{(5)} ,H^{(5)}\right ] \quad \Rightarrow \quad \partial_0 \bar{{
\cal{B}}}_i -  \partial_i\bar{{\cal{B}}}_0 - \epsilon _{ijk}\partial_j \bar{B}_k=0, \nonumber\\
\dot{\Pi}_{0i}^{(5)} &=& -i\left[\Pi_{0i}^{(5)} ,H^{(5)}\right ] \quad \Rightarrow \quad \partial_0 \bar{B}_i 
-  \partial_i \bar{B}_0 + \epsilon _{ijk}\partial_j \bar{{\cal{B}}}_k =0.
\end{eqnarray} 
It is elementary to check that finally we obtain the following simple equations of motion
\begin{eqnarray}
\Box B_{\mu \nu }&=&0,\quad\quad \Box B_\mu =0, \quad \quad \Box\bar{B}_\mu=0, \nonumber\\
\Box{\varphi _1}&=& 0, \quad \quad \Box{\varphi _2}=0, \quad \quad \Box{\cal{B}}_\mu=0,\;\quad \quad 
\Box\bar{{\cal{B}}}_\mu =0, \nonumber\\
\partial \cdot B &=& 0, \quad \quad \partial \cdot\bar{B} = 0, \quad \quad \partial\cdot{\cal{B}} =0, \quad \quad \partial\cdot\bar{{\cal{B}}} = 0,
\end{eqnarray}
from the above Hamiltonians $H^{(4,5)}$.

\end{document}